\documentclass[letterpaper,12pt]{article}
\usepackage{amssymb}
\usepackage{amsthm}
\usepackage{authblk}
\usepackage{setspace}
\usepackage[textwidth=6.5in,textheight=8.25in,centering]{geometry}
\usepackage[pdftex]{graphicx}

\begin{document}

\title{Defect free visible photoluminescence from laser-generated germanium nanoparticles}
\author{Manoj Kumar\thanks{{\scriptsize Present Address: School of Chemical \& Biomedical Engineering, Nanyang Technological University, Singapore.}},
Rajesh Kumar\thanks{{\scriptsize Present Address: National Institute
for Nanotechnology (NINT), University of Alberta, Edmonton, Alberta,
Canada.}} \footnote{{\scriptsize Corresponding Author:
rajesh2@ualberta.ca; Ph: +1 780 709 4513}}, Vivek Kumar and A.K.
Shukla} \affil{Department of Physics, Indian Institute of Technology
Delhi, New Delhi - 110016, India}
\date{\today}

\maketitle
\begin{onehalfspacing}
\begin{abstract}
Origin of room temperature visible photoluminescence (PL) from
defect free germanium (Ge) nanoparticles have been discussed here.
The Ge nanoparticles produced by laser-induced etching technique
show broad visible PL around 2.0 - 2.2 eV at room temperature. Size
dependent PL peak shift in Ge nanoparticles has been explained in
terms of quantum confinement. Theoretical calculations of radiative
lifetime using oscillator strength, which is closely related to the
size of the nanostructures, suggests that the PL is originating from
a radiative recombination process in quantum confined Ge
nanostructures.
\end{abstract}
\end{onehalfspacing}
\vspace{3cm}

PACS: 78.55.-m; 78.67.-n; 78.67.Bf

Keywords: Photoluminescence spectroscopy, Ge nanoparticles, Quantum
confinement.
\newpage
\begin{doublespacing}
\section{Introduction}
Visible electroluminescence and photoluminescence (PL) from
germanium (Ge) nanoparticles have received considerable attention in
the last few years due to its potential applications in
multifunctional electronic and photonic devices [1-4]. The
luminescent Ge nanoparticles can be fabricated by several methods
such as rf-magnetron cosputtering, ion-implantation, dc-sputtering
in a reactive oxygen environment, chemical vapor deposition
technique or by oxidation of Si-Ge alloys with subsequent thermal
annealing to induce the crystallization [2-9]. In most of the
reports, Ge nanoparticles are embedded in a SiO2 matrix and the
observed PL spectrum is broad band around 2.0-2.4 eV [5, 8]. PL band
near 1.83 eV is also reported by some researchers [5]. It is
reported that Ge nanoparticles excited by 488.0 nm emits a visible
PL around 2.2 - 2.3 eV which is independent of size in the 6 - 14 nm
range [4]. However, Y. Maeda and coworkers [2] attributes the
visible PL in the range of 2.0 - 2.3 eV to quantum confinement
effect in Ge nanoparticles. Other reports support the origin of PL
to be defect related because different types of defects in the
$SiO_2$ (i.e. nonbridging oxygen hole centers or NBOHC) can emit
visible PL around 2.3 eV [10]. When Si is partially oxidized, the
energy band gap of nonstoichiometric $SiO_x$ varies in the range of
1.7 - 2.3 eV depending on the oxygen composition [11]. The PL peak
around 2.1 eV may also arise due to some luminescent center at the
interface between Ge nanoparticles and $SiO_2$ matrix [12]. These
reported results indicates that the visible PL can be due to quantum
confinement effect in Ge nanoparticles or due to defects at the
interface of nanoparticles / matrix or in the SiO2 matrix itself and
exact origin of visible PL in the range of 2.0 - 2.3 eV is still
debatable. Although time-resolved PL decay mechanism have been used
to study excitation recombination process for Ge nanoparticles [7].
However it cannot provide any insight about the role of
nanoparticles and nanoparticles-matrix interface electronic states.
Hence the role of Ge nanoparticles need to identify for the observed
visible PL by investigating the effect of quantum confinement by
removing(or minimizing) the effect of defects related effects.

In this Letter, Ge nanoparticles have been prepared by laser induced
etching (LIE) technique to neglect the effect of substrate on the
nanoparticles. The Ge nanoparticles prepared by this method are of
the sizes comparable to Bohr excitonic radius of Ge    (24.3 nm)
[2]. The Ge nanoparticles, on extremely clean (no impurity) surface,
generated in this way show visible PL at room temperature due to
quantum confinement effect. This is supported by the calculation of
radiative life time using the size of nanoparticles, which suggestes
that quantum confinement effect plays the key role in room
temperature visible PL from matrix free laser-generated Ge
nanoparticles.
\section{Experimental Details}
To fabricate Ge nanoparticles, a Ge wafer was immersed in an aqueous
solution of HF acid of 48 \% concentration and irradiated with an
argon-ion laser (514.5 nm). Details of the LIE set-up are given in
the Ref. [13]. Different samples containing Ge nanoparticles of
different sizes have been prepared by varying the etching time in
the range of 35-55 minutes with a fixed laser power of 200 mW. After
irradiation each of the samples was rinsed with ethanol and then
dried before characterization. The PL spectra were recorded with a
computerized spectroscopic system that consists of a SPEX double
monochromator, a HAMAMATSU R943-02 photomultiplier tube. Samples
were excited with photon energy of 2.41 eV from an argon-ion laser
(COHERENT, INNOVA 90-5).
\section{Results and Discussion}
The AFM images in Figs. 1(a) -1(c) shows of time evolution of
laser-etched Ge surfaces during LIE of Ge wafer. One important
observation in Fig. 1 is the formation of smaller Ge nanoparticles
for higher etching times. As etching time increases from 35 minute
to 55 minute, Ge nanoparticles of a few nanometer can be seen in
Fig. 1(c). These results predict that etching time is a controlling
parameter to control the size and size distribution of Ge
nanoparticle. The Ge nanoparticles generated in this method are
formed directly on the Ge wafer itself. In this way the effect of
matrix on PL can be neglected. Variation in the nanoparticle size as
a consequence of increasing etching time can be understood as
follows.

During LIE process, excess holes are generated when n-type Ge wafer
is irradiated with laser light in the presence of HF acid solution.
These laser-generated holes diffuse towards the surface due to an
electric field produced by band bending at the Ge / electrolyte
interface. The capture of holes at the surface leads to removal of
Ge atoms leaving atomic size dip (pore) at the surface. Beale et al.
[14], Lehmann and Gosele [15] have discussed that the pore formation
over the surface change the electric field distribution in such a
way that the holes focus to the bottom of pore. This leads to
preferential etching at the bottom of the pores and etching will be
faster in the substrate direction which leads to pyramidal structure
over the surface. For the longer etching time the height of
structure over the surface would become too long and incident light
will not reach to the base. In this case etching at the top of
structure will dominant and result in a gradual reduction of the
height of the pyramidal structure as well as splitting of the
structure, which leads to reduction of base diameter. As the
nanoparticles over the surface becomes around 4 nm, quantum
confinement effects cause an increase in the band gap [4,16]. Once
the band gap has increased sufficiently, there is no absorption of
photons i.e. there is natural stop for further etching due to band
gap widening [17].

Possibility of quantum confinement effect in laser generated Ge
nanoparticles has been investigated using PL spectroscopy. Fig. 2
shows the room temperature PL spectra from the assembly of Ge
nanoparticles prepared using different etching times. The
comparisons of PL spectra for different irradiation time indicate
that the observed PL ranges from 2.0 - 2.2 eV. Various models
[4,5,8, 10-12] have been proposed to explain this type of observed
PL. The visible PL is attributed either to quantum confinement of
excitions or to be originating from defects at the nanoparticles/
matrix interface or in the matrix itself. However, the quantum
confinement effects can explain the majority of the observed PL
[3,5,18], while in some cases, other mechanism also contributes in
addition to the quantum confinement effect [5]. In the present
study, the nanoparticles are grown (in isolation) directly over the
Ge wafer without any matrix, i.e. the matrix and interface effect is
excluded and should have no effect on observed PL. The LIE process
has involvement of fluorine or hydrogen, so there species such as
$GeF_4$, $GeHF_4$ or $H_2GeF_6$, can form a passivation layer over
the surface and may generate recombination centers for the carriers,
where PL may come due to chemisorbed molecules over the surface
[19]. Thus, it is important to exclude the effect of chemical
species to the see the sole contribution of quantum confinement in
the observed PL. Samples were rinsed with ethanol to remove the etch
products from the surface of nanoparticles before recording the PL
spectra. Thus, the only possible expected origin for visible PL is
the quantum confinement of excitons.

Comparison of PL spectra in Fig. 2(I) for different irradiation
times indicates that as irradiation time increase from 35 to 45
minute there is a strong increase of the PL intensity with an
emission band around 2.1 eV. As irradiation time increases to 55
minutes (Fig. 2(c)) the peak position of band shifts to ~2.2 eV.
Thus, a blueshift of around ~ 1.5 eV is observed for the Ge
nanoparticles in comparison to the bulk Ge. The blueshift of the PL
peak is attributed to the quantum confinement effect in very small
nanoparticles. The broadening of the peak is associated with the
particle size distribution. For 35 minute irradiation time no strong
PL is observed because quantum confinement effect are only
significant when size is comparable to Bohr radius of the excitons.
For irradiation time of 45-55  minute the blueshift of PL is
observed due to enhancement of band gap caused by quantum
confinement, as size is comparable to Bohr excitonic radius [2].

To investigate the range of nanoparticles which contribute
efficiently to visible PL the size and size distribution are
estimated using quantum confinement model proposed by Yorikawa et
al. [20] by the fitting of PL spectra for irradiation time of 45
minute. With the assumption that the each particle has a very sharp
luminescence the PL intensity for the ensemble of particles having a
Gaussian distribution [$D(R_E$)] of sizes can be written as
\begin{equation}
S(E)=C \alpha(E_{exc}-E)D(R_E){\frac{1}{n}}{\frac{R_E}{E-E_g^0}}
\end{equation}
where $R_E$ is the radius of nanoparticles defined by
$R_E=\left(\frac{\beta}{E-E_g^0}\right)^{\frac{1}{n}}$ The $E_g^0$ ~
0.67 eV is the gap energy of bulk Ge and $n$ is 2 for 2D quantum
confinement with $\beta$ as the coupling constant. Two Gaussian
distributions of sizes are used in the Eq. (1) to fit the
experimental data. Doted lines in the Fig. 2(II) display theoretical
PL spectrum for both Gaussian distributions. Continuous lines shown
in the Fig. 2(II) is the superposition of two PL spectrums with two
different size distributions. All fitting parameters and size
distribution used for the line shape fitting of the PL spectrum are
shown in the Table-1. These parameters reveal that the experimental
PL data is the combination of two bands. The band-A at ~1.80 eV is
due to the nanoparticles of average size of 4.5 nm where the size
distribution is in the range of 3.2 - 6 nm. The band-B at ~ 2.10 eV
is due to nanoparticles of average size of 4 nm with size
distribution of 3.2 - 4.8 nm. This is in consonance with the results
of Takagahara et al. [16] where luminescence peak energy of 2.18 eV
corresponds to the quantum dot of average diameter of 4.2 nm. Y.
Maeda [3] and P. K. Giri et al. [5] also observed the similar peaks
in favour of the quantum confinement model. On the basic of observed
results, we conclude that the dominant contribution to PL in the
range of 2.0 - 2.2 eV comes due to confinement in smaller
nanoparticles, and not related to defects at the
nanoparticles/matrix interface or in the $SiO_2$ matrix itself.

The size dependent blue shift of the PL peak position and
significant increase in the PL intensity with decrease in size
indicates that the overall quantum efficiency has also increases
[21]. This attributes to the transition from the indirect to direct
recombination process that causes the increase in oscillator
strength ($f_{osc}$) and it leads to reduction in the radiative
lifetime ($\tau_R$) of carriers from few microseconds to few
nanoseconds [21-23]. Theoretically the $f_{osc}$ and the lifetime
$\tau_R$ for dipole-allowed optical transitions are calculated by
the following simplified equation [24]:
\begin{equation}
\tau_R=\frac{2\pi \varepsilon_0mc^3}{e^2n\omega^2f_{osc}}
\approx1.87\times10^3 \left(\frac{\lambda^2}{nf_{osc}}\right)
\end{equation}
In the above equation we have use the exciton mass ($m = m_e + m_h$)
with $m_e$ = 0.082$m_0$ and $m_h$ = 0.043 $m_0$ after Y. Maeda et al
[2], where $m_0$ is the rest mass of electron. Here$\tau_R$ is in
sec if $\lambda$ is expressed in meter. The effect of the change in
oscillator strength with particle size `$d$' has been addressed by
Khurgin et al. [25] who have expressed the variation of $f_{osc}$
with `$d$' by the following equation:
\begin{equation}
f_{osc}\approx\left[{\frac{\sin\left({\frac{0.86\pi
d}{a}}\right)}{\left[{1-\left({\frac{0.86\pi
d}{a}}\right)^2}\right]\left[\left({\frac{0.86\pi
d}{a}}\right)\right]}}\right]^2
\end{equation}
where $a = 0.565 nm$ is the lattice constant of crystalline Ge. For
the smaller dimensions of the nanoparticles the `$f_{osc}$' can be
approximated by power law of `$d$' as follows
\begin{equation}
f_{osc}\approx d^{-\gamma}
\end{equation}
where $\gamma$ is taken to be 6, Hybertsen and Needels [26] also
suggested the  $\gamma$ = 6, whereas Sanders and Chang proposed a
value of 5 [27]. The variation of the $f_{osc}$ with particle size
$d$ (in the range of 1 -20 nm) for both the values of $\gamma$  is
plotted in Fig. 3. It can be seen that for particle size greater
than 10 nm the $f_{osc}$ is the order of  $10^{-6}$ in both the
cases. However, for particle size below 3 nm (which is the minimum
calculated size for irradiation time of 45 minutes, Table-1) the
fosc increases over several order of magnitude and reaches unity for
particle size $d$ = 1 nm. The Ge nanoparticles of 3 nm  have
$f_{osc}$ equal to $10^{-3}$ and  $\lambda$ as 596 nm [emitted
wavelength for irradiation time 45 minute], which gives radiative
lifetime $\tau_R$ comes to be equal to 160 nm using Eq. (2). The
radiative life time in the range of nanometer suggests that the PL
is not due to defect related effects, where the life times are in
the range of micrometers. The calculated results are in close
agreement with the experimental predicted radiative lifetime for
smaller Ge nanoparticles [5]. Thus the observed PL from
laser-generated Ge nanoparticles are solely due to quantum
confinement effect.
\section{Conclusions}
In summary, PL studies on defect free laser generated Ge
nanoparticles have been done to see the sole effect of quantum
confinement effect on electronic states in low dimensional Ge. Size
dependent blue-shift in PL peak position with decreasing size has
been attributed solely to the quantum confinement effect. The effect
of defect and other possible contributions in visible PL has been
neglected by fabricating the Ge nanoparticles on the Ge wafer
itself. The PL peak position shifts from 2.05 eV to 2.2 eV on
decreasing the size of Ge nanoparticles due to quantum confinement
and is not related to defects at the nanoparticles/matrix interface.
In addition to this oscillator strength, $f_{osc}$ has also been
calculated as a function of particle size, which has been used to
calculate the radiative lifetime. The theoretically calculated
lifetime comes out to be ~ 160 ns as expected from confinement
effect.

\end{doublespacing}

\textbf{Acknowledgements}: The authors are grateful to Prof. V. D.
Vankar (IIT Delhi) for many useful discussions. Authors acknowledge
the financial support from the Department of Science and Technology,
Govt. of India under the project "Linear and nonlinear optical
properties of semiconductor/metal nanoparticles for
optical/electronic devices". Technical support from Mr. N.C.
Nautiyal is also acknowledged. One of the authors (RK) thanks the
financial support from National Research Council (NRC), Canada.
Vivek Kumar thanks University Grants Commission (UGC), India for
providing fellowship

\newpage

\begin{table}
\caption{PL fitting parameters of band A and band B for the
laser-etched Ge nanoparticles used in Eq.(1)} \center

\begin{tabular}{|c|c|c|c|c|c|}
  \hline 
  Band & PL peak position(eV) & L(nm) & $\sigma$(nm) & L$_1$(nm) &
  L$_2$(nm)
  \\ \hline
  A & 1.8 & 4.5 & 0.24 & 3.2 & 6 \\
  B & 2.1 & 4.0 & 0.12 & 3.2 & 4.8 \\
  \hline
\end{tabular}

\end{table}

\newpage
\begin{figure}
\includegraphics[width=15.0cm]{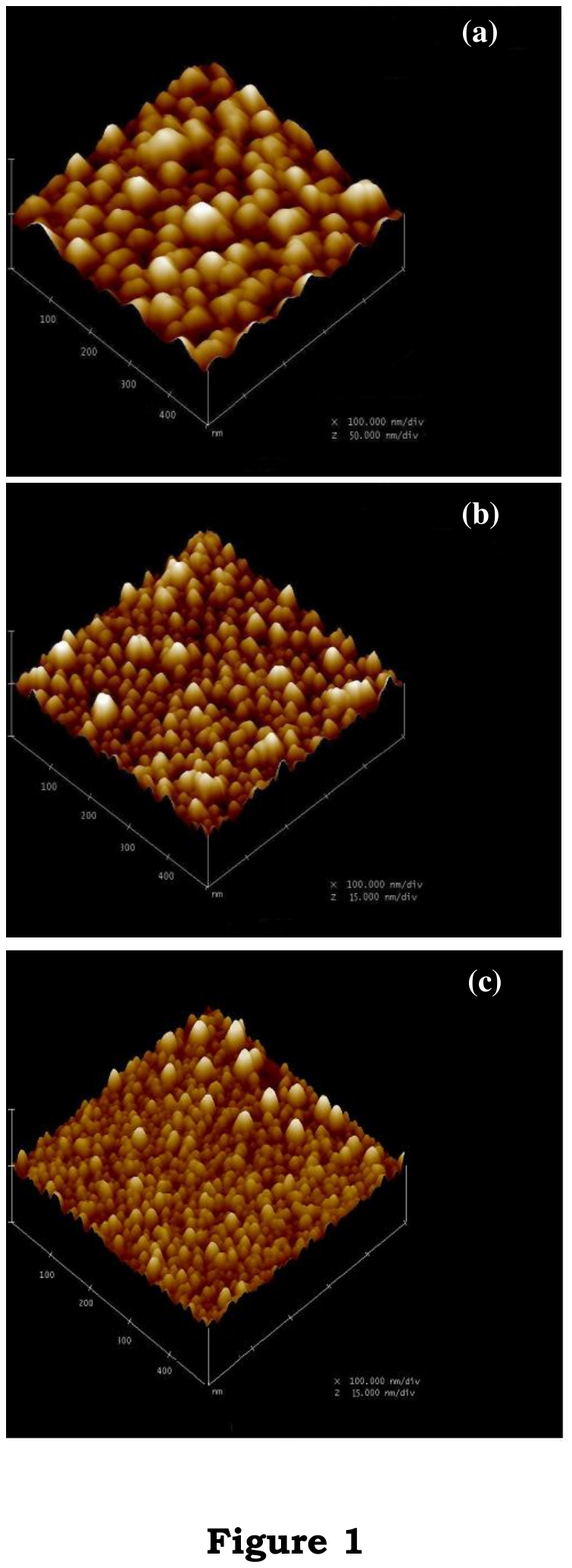}
\caption{AFM images of laser-etched Ge for different etching times
of (a) 35 minute, (b) 45 minute and (c) 55 minute with fixed laser
power of 200 mW.}
\end{figure}

\newpage
\begin{figure}
\includegraphics[width=15.0cm]{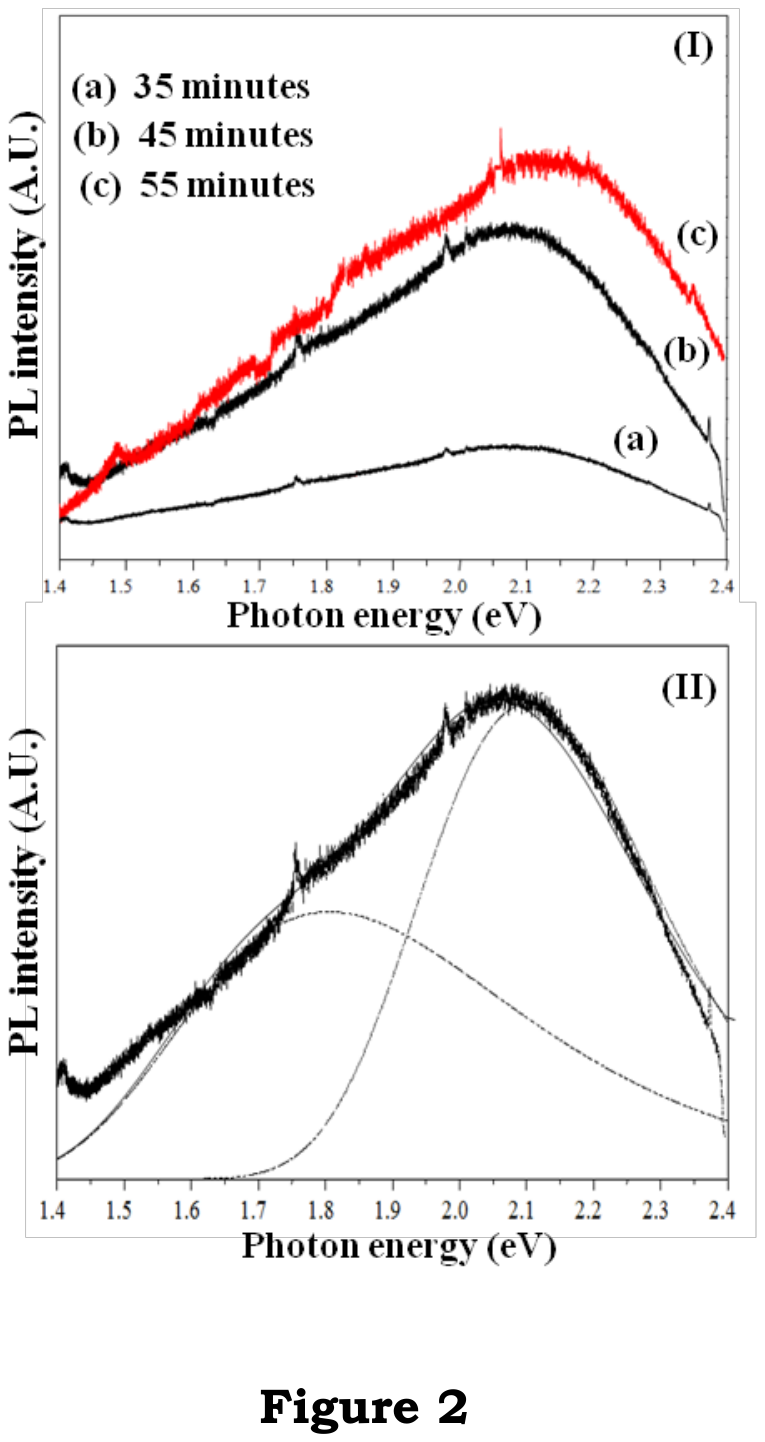}
\caption{(I) PL spectra from the assembly of Ge nanoparticles at
different irradiation times of (a) 35 minute, (b) 45 minute and (c)
55 minute.(II) theoretical fitting of PL spectrum for irradiation
time of 45 minute by confinement model. The theoretically calculated
PL spectra using Eq. (1) is shown by dotted lines.}

\end{figure}

\newpage
\begin{figure}
\includegraphics[width=15.0cm]{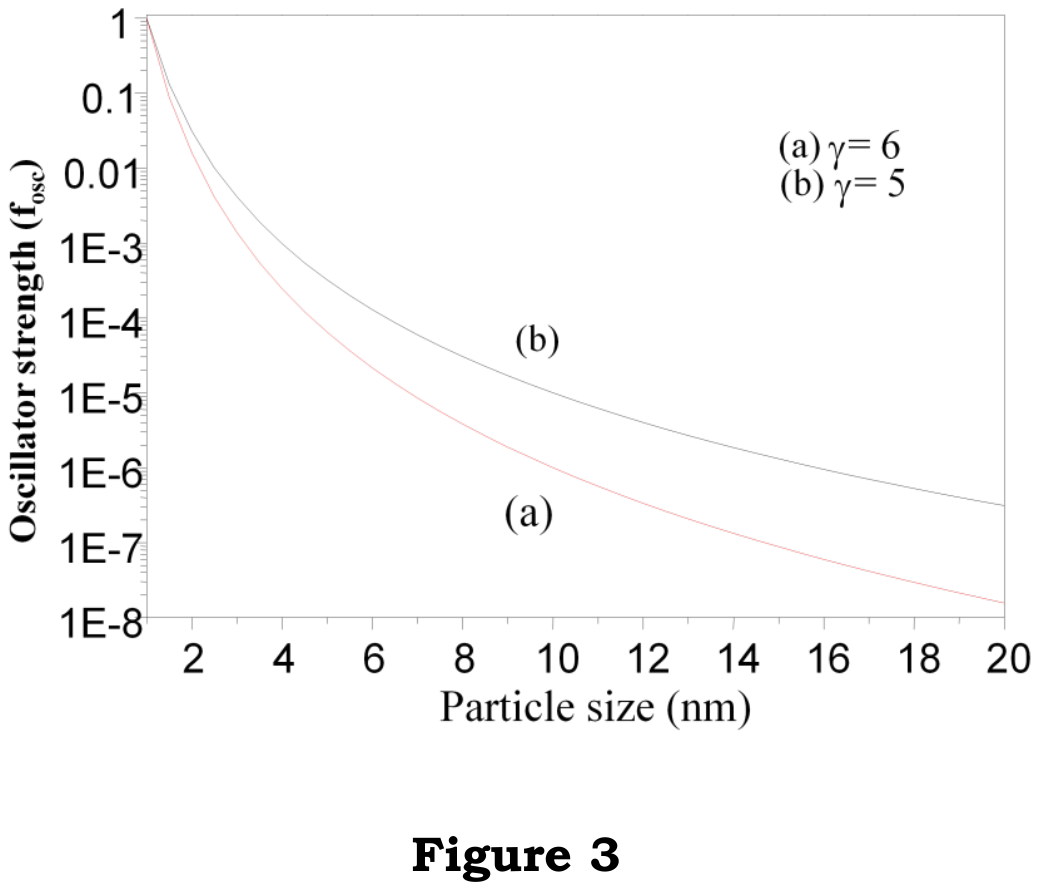}
\caption{Oscillator strength, $f_{osc}$ as a function of particles
size calculated from Eq.(4) for two different values of $\gamma$.}
\end{figure}

\end{document}